\documentclass[aps,prl,a4paper,showpacs,twocolumn]{revtex4}

\input epsf.tex

\usepackage{amssymb}
\usepackage[latin1]{inputenc}
\usepackage{amsmath}
\usepackage{amsfonts}
\usepackage{bm}
\usepackage{graphicx}
\usepackage{epsfig}

\newcommand{\be}{\begin{equation}}
\newcommand{\ee}{\end{equation}}
\newcommand{\ba}{\begin{eqnarray}}
\newcommand{\ea}{\end{eqnarray}}

\begin{document}

\title{Interference of multi-mode photon echoes generated in spatially separated solid-state atomic ensembles} \date{\today} 
\author{M. U. Staudt$^{1}$, M. Afzelius$^{1}$, H. de Riedmatten$^{1}$, S. R. Hastings-Simon$^{1}$, C. Simon$^{1}$, R. Ricken$^{2}$, H. Suche$^{2}$, W. Sohler$^{2}$, N. Gisin$^{1}$}

\author{}
\affiliation{$^{1}$ Group of Applied Physics, University of
Geneva, CH-Geneva, Switzerland\\ $^{2}$ Angewandte Physik,
University of Paderborn, 33095 Paderborn, Germany }

\begin{abstract}

High-visibility interference of photon echoes generated in
spatially separated solid-state atomic ensembles is demonstrated.
The solid state ensembles were LiNbO$_3$ WGs doped with Erbium
ions absorbing at 1.53 $\mu$m. Bright coherent states of light in
several temporal modes (up to 3) are stored and retrieved from the
optical memories using two-pulse photon echoes. The stored and
retrieved optical pulses, when combined at a beam splitter, show
almost perfect interference, which demonstrates both phase
preserving storage and indistinguishability of photon echoes from
separate optical memories. By measuring interference fringes for
different storage times, we also show explicitly that the
visibility is not limited by atomic decoherence. These results are
relevant for novel quantum repeaters architectures with photon
echo based multimode quantum memories.
\end{abstract}

\maketitle
Broad efforts are currently under way to extend quantum
communication to very long distances
\cite{Briegel98,gisin07,Duan01,collins07,simon07,chaneliere05,eisaman05,chou07,boozer07,maunz07}.
Existing quantum communication systems are limited in distance
mainly due to exponential transmission losses. One approach to
overcome this limitation is the implementation of
quantum-repeaters \cite{Briegel98}, which have quantum memories
\cite{chaneliere05,eisaman05,chou07} as the core element. Many
proposed quantum repeater protocols see e. g.
\cite{Duan01,collins07,simon07} exhibit as a common feature the
distribution of entanglement by interference of quantum states of
light stored and released from spatially distant quantum memories
(QMs). The storage must be phase preserving, as
entanglement between remote QMs could not be created otherwise.\\
To obtain reasonable counting rates in quantum communication
systems for distances over 1000 km, some form of multiplexing is
likely to be required \cite{collins07,simon07}. For instance the
protocol of Ref. \cite{simon07} predicts a speed up in the
entanglement generation rate by taking advantage of the storage of
multiple distinguishable temporal modes (multi-modes) in a single
quantum memory. Techniques based on photon echoes
\cite{kurnit64,moss82} seem well adapted for the storage of
multi-modes, as photons absorbed at different times are emitted at
different times. Multi-mode storage of classical pulses has been
successfully implemented in swept-carrier photon-echo experiments,
where up to 1760 modes have been stored and retrieved
\cite{lin95}. The storage of single photons with high efficiency
using a photon-echo type scheme is in principle possible with a
technique based on Controlled Reversible Inhomogeneous Broadening
(CRIB)\cite{mois01,nilsson06,alex06}, where inhomogeneous
rephasing is triggered by an external electric field instead of a
strong optical pulse. \\Atomic ensembles in the solid state using
rare-earth-ion-doped materials seem promising for the
implementation of a quantum memory, owing to the absence of atomic
diffusion and the long coherence times that are possible for
optical and hyperfine transitions \cite{longdell05,sun02}. For the
realization of multi-mode QMs using photon echo techniques, long
optical coherence times are essential in order to achieve
sufficient storage times and high efficiency storage for several
temporal modes \cite{simon07}. Moreover, there is a wide range of
wavelengths available using different rare-earth-ions. In
particular, the Erbium ion has a transition around 1530 nm, where
the transmission loss in optical fibers is minimal. First steps
towards photonic quantum storage in rare-earth-ion doped materials
have been demonstrated using Electromagnetically Induced
Transparency \cite{longdell05} and photon echoes approaches
\cite{alex06,staudt07}. Several experiments have shown phase
preserving storage of bright pulses in a single optical memory
with photon echo techniques \cite{staudt07,arend93,alex07}. \\In
this paper we address the issue of phase preservation in the
storage and retrieval of bright coherent states in multiple
temporal modes in two different solid state optical memories,
based on the photon echo technique. The atomic ensembles,
separated by 7 cm, are implemented with Erbium ions doped into a
LiNbO$_3$ waveguide (WG). We also study the influence of atomic
decoherence on phase preservation by varying the storage time. The
implementation of CRIB requires an efficient three-level lambda
system, which has not been demonstrated in Erbium so far. However,
the phase preservation for a CRIB based quantum memory can be
investigated in two-pulse photon echo based memory using bright
pulses, as phase properties of the storage material will not
change.\\To investigate the phase preservation, we use first order
interference of photon echoes generated in two Erbium-doped
LiNbO$_3$ WGs placed in the arms of a balanced Mach-Zehnder
interferometer (see Fig. 1). Two-pulse photon echoes are generated
in the two ensembles by coherent excitation using two bright
pulses, and the photon echoes created in each arm interfere at the
second beam splitter. The interference fringe visibility is used
as a measure of the phase coherence of the memories.\\In a
two-pulse photon echo experiment using an inhomogeneously
broadened two-level atomic system, a first pulse of area $\Theta$
(called data pulse) brings the atoms into a coherent superposition
of ground and excited state. A macroscopic dipole moment is
produced and decays due to inhomogeneous dephasing. A second
pulse, ideally a $\pi$ pulse (called read pulse) a time
\textit{t$_{12}$} after the data pulse will lead to rephasing, and
after a time 2\textit{t$_{12}$} a macroscopic dipole moment is
established producing a photon echo. The two-pulse photon echo can
be seen as a storage and retrieval of the data pulse, which can
also be a sequence of pulses. The time \textit{$t_s=2t_{12}$} is
the storage time.
\begin{center}
\begin{figure}
\epsfxsize=0.5\textwidth \epsfbox{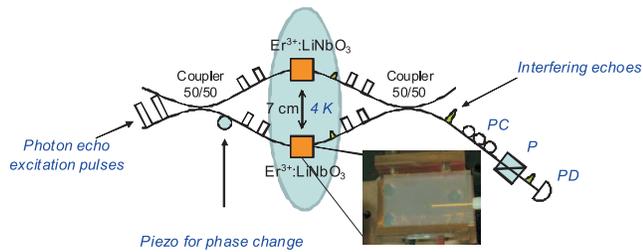} \caption{Experimental
setup: two Er$^{3+}$:LiNbO$_3$ WGs (a photo of one is shown in the
inset) are placed in the arms of a fiber-optic interferometer in a
region where the interferometer is at 4 K. While both arms have
the same length (2.63 m), in one arm the fiber is coiled partially
around a piezo-electric element, allowing for controlled phase
shifts. The excitation light pulses are sent through the
interferometer. The generated echoes interfere at the second
coupler. In order to project the polarizations onto one axis we
used a polarization controller (PC), a polarizer (P) and a photo
detector (PD)}\label{fig4}
\end{figure}
\end{center}
The memories used were two Erbium doped LiNbO$_3$ crystals with
single-mode Ti-indiffused optical WGs (for more details see Refs
\cite{staudt07,Baumann97}). The atoms were excited on the
$^4I_{15/2}\rightarrow^4I_{13/2}$ transition at a wavelength of
1532 nm. This transition is inhomogeneously broadened to 250 GHz
due to slightly different local environments seen by each Erbium
ion \cite{sun02}. The WGs had a length of about 20 mm, where 10 mm
were doped. WG II had two times higher Erbium doping concentration
than WG I (WG I: $4*10^{19}/cm^3$ surface concentration before
indiffusion), resulting in a higher absorption in WG II. The
optical coherence time is limited by magnetic spin interactions
between Erbium ions, but can be considerably increased by applying
an external magnetic field \cite{boettger06}. A constant magnetic
field of about 0.2 Tesla was applied over WG I and a variable
field using a supra-conducting magnet was applied over WG II, both
fields being parallel to the C$_{3}$ axis of the LiNbO$_3$
crystal. The different magnetic fields and Erbium doping
concentrations resulted in different optical coherence times,
T$_2=18~\mu s$ in WG I and T$_2=6~\mu s$ in WG II (at around 0.5
Tesla) at a temperature of about 3 K.\\The fiber-optic
interferometer was placed within a pulse-tube cooler (Vericold).
Both arms were installed across a temperature gradient of 300 K,
as the couplers had to be placed at ambient temperature. One
Er$^{3+}$-doped LiNbO$_3$ WG was mounted in each arm on the 4
K-plate, separated spatially by 7 cm. The photon echo excitation
light pulses had durations of $t_{pulse}$=15 ns and the pulse
sequence was repeated at a frequency of 13.5 Hz. This frequency
was found to minimize the phase noise of the interferometer due to
vibrations in the pulse-tube cooler. The light pulses were created
by gating an external-cavity cw diode laser (Nettest Tunics Plus)
with an intensity modulator, before amplification by an EDFA
(Erbium Doped Fiber Amplifier). An additional acousto-optical
modulator (AA opto-electronics) between the optical amplifier and
the input of the pulse-tube cooler, which opened only for the
series of pulses, helped to suppress light for all other times
even further, thus avoiding spectral hole burning by the EDFA. The
input peak powers were in the order of 60 mW (300 mW) for the data
(read) pulses and the released echoes were in the order of a few
$\mu$W due to the efficiency of the echo process (about 1 $\%$)
and losses in the interferometer.\\In order to obtain high
visibility interference fringes with the retrieved photon echoes,
two criteria must be fulfilled. (1) The storage must be phase
preserving (i.e. the two echoes must have a fixed phase relation)
and (2) the photon echoes must be indistinguishable in order to
avoid which-path information. In order to satisfy the second
criterion, the echoes from the two ensembles must be detected in
the same spatial, polarization and spectral/temporal mode.
Moreover, the intensities of the two echoes in front of the
detector must be the same. Spatial indistinguishability is ensured
by the fact that the interferometer is completely single-mode,
owing to the use of single mode fibers and WGs. The polarization
of the photon echoes could in principle be adjusted by using
polarization maintaining fibers or by inserting polarization
controllers inside the interferometer. For technical reasons, we
chose instead to project the two photon echoes on a common
polarization axis, by inserting a polarizer and a polarization
controller before the photo-detector (see Fig. 1). Since the
generation of photon echoes is a non-linear process, the
spectral/temporal modes of the echoes depend strongly on the
intensity of the excitation pulses and on the optical depth
\cite{wang99}. Since the two WGs have different absorption depths,
the temporal shape could therefore be adjusted by tuning the
wavelength of the excitation laser within the absorption profile.
Finally, to equalize the intensity of the echoes, we used an
ultra-high precision translation stage system (anp, attocube
systems) to change the in- and out coupling powers. Moreover, by
adjusting the magnetic field for WG II one can change the optical
coherence time, and thus change the amplitude of the photon echo.
\\In order to characterize the phase noise of the interferometer,
we measured interference fringes using a cw-off-resonant laser at
$\lambda=1550$ nm. While scanning the phase as the pulse-tube was
cooling a visibility close to $92\%$ was obtained, limited by
phase noise introduced by the vibrations of the cooling system. By
shutting down the cooling system for a short time, we obtained a
visibility close to $100\%$. Thus for all photon echo
measurements, which had to be carried out below 4K, a visibility
of 92$\%$ was an upper technical limit.
\begin{center}
\begin{figure}
\epsfxsize=0.48\textwidth \epsfbox{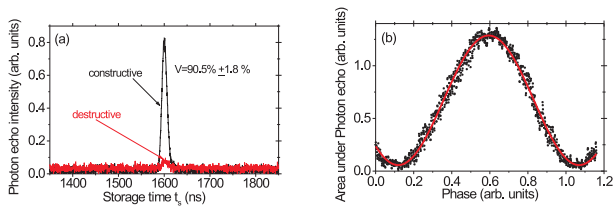} \caption{(a)
Intensity of photon echoes for constructive (black curve) and
destructive (red curve) interference. (b) Interference fringe: The
area under the interfering photon echoes generated in two
spatially separated WGs is shown as a function of phase
difference. The storage time was set to $1.6~\mu$s. Each point is
averaged over 10 echoes and the measurement of a whole fringe
takes about 10 minutes. For this particular fringe, a visibility
of V=91.5$\%$ is reached (averaging over many measurements gives a
visibility of 90.5 $\%$), limited by phase noise caused by
vibrations in the cooling system.}\label{fig2}
\end{figure}
\end{center}
We first discuss the experiments carried out on the storage of a
single temporal mode. In these experiments the storage time was
fixed at $1.6~\mu$s. Fig.~\ref{fig2} a shows interfering photon
echoes for constructive and destructive interference. By scanning
the phase difference of the interferometer continuously,
interference fringes were obtained, an example is shown in
Fig.~\ref{fig2} b. A visibility of 90.5$\%$ ($\pm1.8\%$) was
obtained by averaging over many measurements, which is within the
limit set by the intrinsic phase noise due to the cooling system.
Therefore we can conclude that the storage of the optical pulses
in the solid state ensembles is fully phase preserving within the
error bars. Moreover the two paths of the interferometer cannot be
distinguished in any of the spatial, polarization or
spectral/temporal modes. Note that high visibilities were
achieved, even though the WGs had different physical properties,
particularly in their absorption coefficients and optical
coherence times.
\begin{center}
\begin{figure}
\epsfxsize=0.48\textwidth \epsfbox{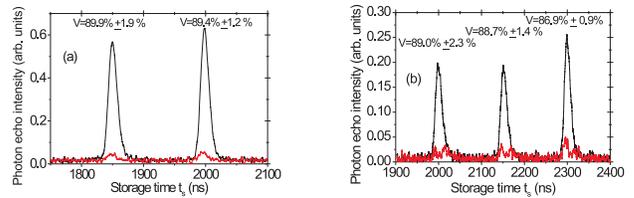} \caption{Multi-mode
storage: Constructive and destructive photon echo interference is
shown for a storage time of $2~\mu$s for two and $2.3~\mu$s for
three stored modes (values for the longest stored mode). Phase
coherence is in the two cases preserved to a very high degree as
indicated by high fringe visibilities.} \label{fig3}
\end{figure}
\end{center}
We now turn to the investigation of the storage of multiple modes.
The data then contains several pulses separated by 150 ns which
coherently excite the atoms. If the intensity of the pulses is
well below saturation, each pulse independently generates an echo
triggered by the strong read pulse. In Fig.~\ref{fig3} we show
examples of interfering photon echo signals for two and three
modes, both for destructive and constructive interference. The
visibility is measured for each mode separately from interference
fringes (as in Fig. 2b) and is an average over several independent
measurements. The average visibility is only slightly reduced from
90.5$\%$ for the case of the storage of one mode to 86.9$\%$ for
the lowest value out of the three stored modes. Thus, we show that
high-visibility storage of multiple modes in independent
solid-state optical memories is possible. In the case of only one
stored mode all parameters (intensity, pulse length, coupling
etc.) were optimized in order to obtain a strong photon echo
signal from both WGs. For the case of two and three stored modes
optimization is experimentally more demanding, as one has to work
in a regime where each of the data pulse areas $\Theta$ has to be
much smaller than $\pi/2$. Otherwise echo distortion and
additional undesired multi-pulse echoes, that might interfere with
the echoes of the data pulses, will occur. Moreover, the issues
related to different optical depths in the two crystals mentioned
above \cite{wang99} are even more critical in the case of multiple
modes, since optimization of one mode does not guarantee the
indistinguishability of the other modes. These issues together
with the detector sensitivity were the main limitations for the
number of modes that could be stored. Overcoming these technical
issues, the enlargement of the storage capacity should be
possible. By implementing a CRIB-type scheme, storage of even more
modes should be possible \cite{simon07}.\\Finally, we study the
effect of atomic decoherence on the emitted photon echoes. The
optical coherence in solid state systems is perturbed by
interaction with the environment, leading to atomic decoherence.
In Erbium doped LiNbO$_3$ this dephasing arises from interactions
between the large Er$^{3+}$ magnetic moment and fluctuating local
magnetic fields due to changing magnetic spins on neighboring
Er-ions, so-called spin flips \cite{sun02}. The atomic decoherence
will cause an exponential decay of the two-pulse photon echo
signal, as exemplified for WG II in Fig. \ref{fig4}. From this
decay of the photon echo signal, a T$_2$ of 6~$\mu$s was
measured.\\In an optical memory, ideally the phase should be
preserved independently of the storage time. At first sight this
might seem to be a contradiction due to the unavoidable atomic
decoherence. To test the influence of atomic decoherence on the
visibility, we measured interference fringes as a function of the
storage time (0.8~$\mu$s to 5.6~$\mu$s, as shown in Fig.
\ref{fig4}). We found that although the photon echo amplitude
decreases strongly due to atomic decoherence in this time period,
the visibility is unaffected and remains at 90.5$\%$. This can be
interpreted as follows: the photon echo signal is a coherent
collective signal, which is due to interference of the photon
emission amplitudes from the whole atomic ensemble
\cite{abella66,dicke54}.
\begin{figure}[th]
\includegraphics[width=0.4\textwidth,angle=0]{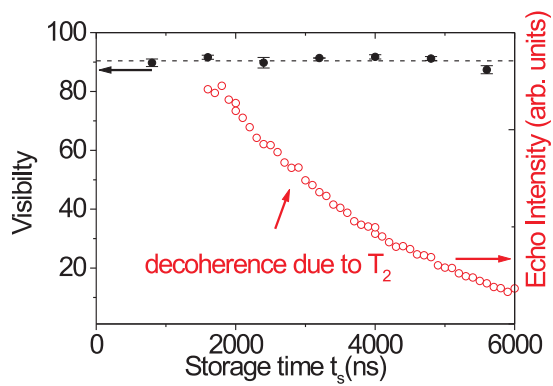}
\vspace{-0cm} \caption{Interference fringe visibility (filled
circles) is shown as a function of the memory storage time. Atomic
decoherence strongly acts on the amplitude of the echo signal, as
shown here for WG II (open circles), but leaves the visibility
unaffected. The dotted line shows the average visibility of
90.5$\%$. For each storage time we readjusted all experimental
parameters, as the optical coherence times are different in the
two WGs.}  \label{fig4}
\end{figure}
Its intensity is proportional to (\textit{N-N'})$^2$, where
\textit{N} is the total number of atoms and \textit{N}' is the
number of decohered atoms. Atoms having lost their coherence due
to interaction with the environment do not contribute to the
collective photon echo emission. Besides the collective coherent
emission in the forward mode, there is also an incoherent emission
from all atoms, which would be there even without a data pulse,
and whose total intensity in the forward spatial mode scales only
with \textit{N}. Therefore, the ratio of the collective photon
echo emission to the incoherent background emission is
(\textit{N-N'})$^2$/\textit{N}. Since \textit{N} is a very large
number (\textit{N}$\sim 10^{8}$) we detect only the coherent
collective part of the signal, even if \textit{N}' is comparable
to \textit{N}. Thus it is within expectation that the decay in
photon echo amplitude has no effect on visibility. This holds as
long as the background noise from other sources than the photon
echo process itself is much smaller than the photon echo
amplitude, as is the case in this experiment. Note in this context
that CRIB is also a collective process. Thus the atomic
decoherence should not influence the phase preservation for a
CRIB-type memory either.
\\Note that in the context of quantum communication, only the
photons re-emitted from the memories and detected contribute to
the signal. Hence, in our case, atomic decoherence which acts only
as a loss will affect the efficiency of a given protocol, but not
the effective fidelity. In that sense, the visibility of the
interference fringe can be considered as a measure of the
effective fidelity of the storage and retrieval in the memory.
\\In conclusion we show that phase coherence is preserved to a high
degree in the storage and retrieval of light pulses in spatially
separated solid state atomic ensembles using photon echo
techniques.
 This phase coherence is revealed in
interference patterns that show a visibility of 90.5$\%$
($\pm1.8\%$), close to the technical limit of 92$\%$ set by phase
noise caused by the cooling system. Extending our studies to three
stored temporal modes only leads to a slight decrease in
visibility, mainly due to the nonlinear character of the photon
echo storage process. Atomic decoherence due to interactions with
the environment does not affect the visibility, i.e. it causes
only a loss of signal amplitude, but not a loss of phase coherence
of the re-emitted optical pulses. This is explained by the
collective character of the photon echo process. These results are
interesting in the framework of proposed novel quantum-repeater
architectures, where storage and retrieval with high phase
coherence preservation and in multiple modes is highly
advantageous.\\We would like to thank G. Fernandez, M. Legr\'{e}
and C. Barreiro for assistance and W.Tittel for discussions. We
are grateful for support by the Swiss NCCR Quantum Photonics and
by the EU Integrated Project Qubit Applications.

\end{document}